\input harvmac
\noblackbox

\Title{\vbox{\baselineskip12pt\hbox{hep-th/0511215}\hbox{SU-ITP-05/32}\hbox{SLAC-PUB-11573}}}
{\vbox{{
\centerline{New Attractors}
\centerline{}
\centerline{and}
\centerline{}
\centerline{Area Codes}
\centerline{}
}}}

\centerline{Alexander Giryavets\footnote{*}{On leave from Steklov Mathematical Institute, Moscow, Russia}\footnote{$ $}{giryav@stanford.edu}}
\medskip
\centerline{Department of Physics and SLAC}
\centerline{Stanford University}
\centerline{Stanford, CA 94305/94309 USA}

$\;$

$\;$

\noindent
In this note we give multiple examples of the recently proposed
New Attractors describing supersymmetric flux vacua and non-supersymmetric
extremal black holes in IIB string theory.
Examples of non-supersymmetric extremal black hole attractors arise
on a hypersurface in $WP^{4}_{1,1,1,1,2}$.
For flux vacua on the orientifold of the same hypersurface existence of multiple basins
of attraction is established. It is explained that certain fluxes may give rise to multiple
supersymmetric flux vacua in a finite region on moduli space, say at the Landau-Ginzburg point
and close to conifold point. This suggests the existence of multiple basins for flux vacua
and domain walls in the landscape for a fixed flux and at interior points in moduli space.

$\;$

\Date{November 2005}

\noblackbox

\def\IP{\relax{\rm I\kern-.18em P}}

\lref\FKS{S. Ferrara, R. Kallosh and A. Strominger,
``$N = 2$ extremal black holes,'' Phys. Rev. D52 (1996) 5412,
hep-th/9508072.}

\lref\Strominger{A. Strominger,
``Macroscopic entropy of $N = 2$ extremal black holes,'' Phys. Lett. B383 (1996) 39,
hep-th/9602111.}

\lref\FKone{S. Ferrara and R. Kallosh,
``Supersymmetry and attractors,'' Phys. Rev. D54 (1996) 1514,
hep-th/9602136.}

\lref\FKtwo{S. Ferrara and R. Kallosh,
``Universality of supersymmetric attractors,'' Phys. Rev. D54 (1996) 1525,
hep-th/9603090.}

\lref\Moore{G. Moore, ``Les Houches Lectures on strings and arithmetic,''
hep-th/0401049.}

\lref\Trivedi{K. Goldstein, N. Iizuka, R. Jena and S. Trivedi,
``Non-supersymetric attractors,'' hep-th/0507096.}

\lref\TripathyTrivedi{P. Tripathy and S. Trivedi,
``Non-supersymetric attractors in string theory,'' hep-th/0511117.}

\lref\Sen{A. Sen, ``Black hole entropy function and the attractor and the attractor
mechanism in higher derivative gravity,'' hep-th/0506177.}

\lref\GKP{S. Giddings, S. Kachru, and J. Polchinski, ``Hierarchies
from Fluxes in String Compactifications,'' Phys. Rev. {\bf D66}
(2002) 106006, hep-th/0105097.}

\lref\KKLT{S. Kachru, R. Kallosh, A. Linde and S. Trivedi, ``de Sitter Vacua
in String Theory,'' Phys. Rev. {\bf D68} (2003) 046005, hep-th/0301240.}

\lref\Kalloshone{R. Kallosh, ``Flux vacua as supersymmetric attractors,'' hep-th/0509112.}

\lref\Kalloshtwo{R. Kallosh, ``New Attractors,'' hep-th/0510024.}

\lref\Denef{F. Denef and M. Douglas, ``Distributions of flux vacua,'' JHEP 0405 (2004) 072,
hep-th/0404116.}

\lref\Eva{E. Silverstein, ``TASI/PiTP/ISS Lectures on Moduli and Microphysics,'' hep-th/0405068.}

\lref\Grana{M. Grana, ``Flux compactifications in string theory: a comprehensive review,''
hep-th/0509003.}

\lref\GKTT{A. Giryavets, S. Kachru, P. Tripathy and S. Trivedi,
``Flux Compactifications on Calabi-Yau Threefolds,''
JHEP {\bf 0404} (2004) 003, hep-th/0312104.}

\lref\GKT{A. Giryavets, S. Kachru and P. Tripathy,
``On the taxonomy of flux vacua,''
JHEP {\bf 0408} (2004) 002, hep-th/0404243.}

\lref\DGKT{O. DeWolfe, A. Giryavets, S. Kachru and W. Taylor,
``Enumerating flux vacua with enhanced symmetries,''
JHEP {\bf 0502} (2005) 037, hep-th/0411061.}

\lref\Dewolfe{O. DeWolfe,
``Enhanced symmetries in multiparameter flux vacua,'' hep-th/0506245.}

\lref\FGK{S. Ferrara, G. Gibbons and R. Kallosh,
``Black holes and critical points in moduli space,''
Nucl. Phys. B500 (1997) 75, hep-th/9702103.}

\lref\DGKTtwoA{O. DeWolfe, A. Giryavets, S. Kachru and W. Taylor,
``Type IIA moduli stabilization,''
JHEP {\bf 0507} (2005) 066, hep-th/0505160.}

\lref\CDKV{A. Ceresole, G. Dall'Agata, R. Kallosh and A. Van Proeyen,
``Hypermultiplets, domain walls and supersymmetric attractors,''
Phys.Rev. D64 (2001) 104006, hep-th/0104056.}

\font\cmss=cmss10 \font\cmsss=cmss10 at 7pt

\def\IC{\relax\hbox{$\inbar\kern-.3em{\rm C}$}}
\def\IR{\relax{\rm I\kern-.18em R}}
\def\Z{\relax\ifmmode\mathchoice
{\hbox{\cmss Z\kern-.4em Z}}{\hbox{\cmss Z\kern-.4em Z}}
{\lower.9pt\hbox{\cmsss Z\kern-.4em Z}}
{\lower1.2pt\hbox{\cmsss Z\kern-.4em Z}}\else{\cmss Z\kern-.4emZ}\fi}

\newsec{Introduction}

Attractors in physics have always been a subject of intensive research.
Among the examples are the supersymmetric extremal black hole attractors found in
\FKS, \Strominger, \FKone, \FKtwo\ (for a nice review see \Moore).

Another recent development, seemingly unrelated to black hole attractors, which
attracts a lot of attention is flux compactification
\GKP, \KKLT, \DGKTtwoA. This development
is reviewed nicely in \Eva\ and \Grana\ (see adequate bibliography here as well).

To be really catchy this note concerns both.
Equations in both cases are pretty similar
but the relation was unclear until recently, when the New Attractors were introduced
in papers \Kalloshone\ and \Kalloshtwo\ (using \Denef). These equations describe
an analog of the supersymmetric extremal black hole attractor equations for supersymmetric
flux vacua (and actually for non-supersymmetric ones too).

Here we choose a particular Calabi-Yau model (a hypersurface in $WP^{4}_{1,1,1,1,2}$)
and show that the known examples of supersymmetric
flux vacua (both Minkowski and AdS) of IIB string theory satisfy the New Attractor equations.
Explicit examples of supersymmetric flux vacua on Calabi-Yau three-folds
were given in \GKTT, \GKT, \DGKT, \Dewolfe\ (see \Grana\ for complete bibliography).
The particular classes of vacua checked here are at the Landau-Ginzburg point
and in vicinity of the conifold point of the model,
which are both in a finite region on the moduli space.

One observation of this note is the existence of area codes. Certain fluxes
give rise to multiple vacua on moduli space. Examples are given of such phenomena
where, for fixed $F_{(3)}$ and $H_{(3)}$ flux, one supersymmetric flux vacuum is at
the Landau-Ginzburg point while other is very close to the conifold.
The values of the AdS cosmological constants in this case are close, but not the same.
This suggests the existence of domain walls in the landscape (for further details see \CDKV).

Other observations we make have to do with the non-supersymmetric extremal black holes in IIB
string theory introduced in \FGK. Interesting recent developments on this subject appear
in \Trivedi, \Sen\ and \TripathyTrivedi. In \Kalloshtwo\
New Attractor equations for these non-supersymmetric black holes were proposed.

We give explicit examples of non-supersymmetric extremal black holes attractors
and it is checked that the New Attractor equations for non-supersymmetric
black holes work for them.

\newsec{Review of New Attractors for flux vacua}

In papers \Kalloshone, \Kalloshtwo\ the New Attractor equations for
flux vacua were proposed. The following attractor equations for $N = 1$
supersymetric flux vacua of IIB string theory were found:

\eqn\atreq{
\pmatrix{h \cr f \cr}
=\pmatrix{ 2{\rm Re}(Z {\overline \Pi}) \cr 2{\rm Re}(Z{\overline \tau}{\overline \Pi}) \cr}
+\pmatrix{ 2{\rm Re}(Z^{{\underline 0} I} {\overline {D_{I}\Pi}}) \cr 2{\rm Re}(\tau {Z}^{{\underline 0} I}{\overline {D_{I}\Pi}}) \cr }
}
where $\tau$ is the axio-dilaton; $f$ and $h$ are the magnetic and electric charges associated
with the RR 3-form flux $F_{(3)}$ and NSNS 3-form flux $H_{(3)}$ of IIB respectively.

$\Pi$ is the covariantly holomorphic section of flat symplectic bundle of dimension
$2h_{2,1} + 2$ over the K\"ahler manifold which obeys the symplectic constraint
\eqn\constraint{
\Pi^{\dagger}\cdot\Sigma\cdot {\Pi} = i
}

The superpotential is defined to be
\eqn\Z{
Z = (f-\tau h)\cdot\Sigma\cdot\Pi
}
while the mass matrix components are
\eqn\ZI{
Z^{{\underline 0} I} = D^{\underline 0}D^{I}Z
}
with the flat derivatives are defined as $D^{I} = e_{i}^{I}D^{i}$ for the symplectic bundle
and $D^{\underline 0} = e_{\tau}^{\underline 0}D^{\tau}$ for the axio-dilaton.

\newsec{Review of supersymmetric flux vacua}

As the properties of flux superpotentials on Calabi-Yau orientifolds
in type IIB string theory have been reviewed  many
times, we will be brief.  Our conventions are those of \GKT.

Consider a Calabi-Yau threefold $M$ with $h_{2,1}$ complex
structure deformations. Choose a symplectic basis $\{A^a, B_b \}$
for the $b_3 = 2 h_{2,1} + 2$ three-cycles, $a,b = 1, \ldots, h_{2,1}+ 1$,
with dual cohomology elements $\alpha_a$, $\beta^b$ such that:
\eqn\cyclebasis{
\int_{A^a} \alpha_b = \delta^a_b \,, \quad \quad \int_{B_b} \beta^a =
- \delta_b^a \,, \quad \quad \int_{M} \alpha_a \wedge \beta^b = \delta_a^b.
}
Fixing a normalization for the unique holomorphic three-form
$\Omega$, let us assemble the periods $z^a \equiv \int_{A^a} \Omega$,
${\cal G}_b \equiv \int_{B_b} \Omega$ into a $b_3$-vector $\Pi(z) \equiv
({\cal G}_b, z^a)$.
The $z^a$ are taken as projective coordinates on
the complex structure moduli space, with ${\cal G}_b = \partial_b{\cal G}(z)$.
The K\"ahler potential ${\cal K}$ for the $z^a$ as well as
the axio-dilaton $\tau \equiv C_0 + i e^{-\varphi}$ is
\eqn\kahler{
{\cal K} = -\log (i \int_{M} \Omega \wedge \overline\Omega) - \log (-i (\tau - \bar\tau))
= - \log(-i \Pi^\dagger \cdot \Sigma \cdot \Pi) - \log(-i(\tau - \bar\tau)) \,,
}
where $\Sigma$ is the symplectic matrix $\Sigma \equiv \pmatrix{\;0\;\;\; 1 \cr -1\;\; 0}$.
The axio-dilaton and complex structure moduli take values in the moduli space ${\cal M}$;
a correct global description of the moduli space requires to identify points in ${\cal M}$
related by modular symmetries.

Now consider nonzero fluxes of the RR and NSNS 3-form field
strengths $F_{(3)}$ and $H_{(3)}$ over these three cycles, defining the
integer-valued $b_3$-vectors $f$ and $h$ via
\eqn\fluxes{
F_{(3)} = - (2 \pi)^2 \alpha'(f_a\, \alpha_a + f_{a +h_{2,1}+1}\, \beta^a)
\,, \quad H_{(3)} = - (2 \pi)^2 \alpha'(h_a \, \alpha_a + h_{a+h_{2,1}+1} \,\beta^a) \,.
}
These fluxes induce a superpotential
for the complex structure moduli as well as the axio-dilaton:
\eqn\gvw{ W = \int_{M} G_{(3)} \wedge \Omega(z) = (2\pi)^2\alpha'\,(f - \tau h) \cdot \Pi(z) \,, }
where $G_{(3)} \equiv F_{(3)} - \tau H_{(3)}$.

We will be interested exclusively in vacua satisfying the F-flatness conditions:
\eqn\origfflat{
D_\tau W = D_a W = 0 \,,
}
where
$D_a W \equiv \partial_a W + W \partial_a {\cal K}$,
and we have allowed $a$ to run only over the $h_{2,1}$ inhomogeneous coordinates.
This is alternatively
\eqn\fflat{
(f - \bar\tau h) \cdot \Pi(z) = (f - \tau h) \cdot (\partial_a \Pi + \Pi \partial_a {\cal K}) = 0 \,.
}
These conditions force the complex structure to align such that  the $(3,0)$ and $(1,2)$
parts of the fluxes vanish, leaving the fluxes ``imaginary self-dual," $*_6 G_{(3)} = i G_{(3)}$.

The fluxes also induce a contribution to the total D3-brane charge
\eqn\nflux{
N_{\rm flux} = {1 \over (2 \pi)^4 (\alpha')^2} \int_{M}  F_{(3)} \wedge H_{(3)}  = f \cdot \Sigma \cdot h \,.
}
In the rest of the paper, we will set $(2\pi)^2 \alpha^\prime = 1$ for convenience.
For vacua satisfying \origfflat, the physical dilaton condition Im $\tau > 0$ implies that $N_{\rm flux} > 0$.
As the total charge on a compact manifold must vanish, sources of negative D3-charge must be present as well.
For a given IIB orientifold compactification, a fixed amount of negative charge is induced by
the orientifolds, leading to an effective bound on $N_{\rm flux}$:
\eqn\fluxbound{
N_{\rm flux} \leq L \,,
}
where, for instance in a IIB orientifold arising as a limit of a
fourfold compactification of F-theory, $L$ can be computed from
the Euler character of the fourfold.  Although the number of imaginary-self
dual flux vacua is infinite, the set satisfying \fluxbound\ for fixed $L$ is
in general finite.

In the absence of fluxes, a symmetry group ${\cal G}= SL(2,Z)_\tau \times \Gamma$ acts
on the moduli space ${\cal M}$, where $SL(2,Z)_\tau$ is the S-duality of type IIB string
theory and $\Gamma$ is the modular group of the complex structure moduli space.
Points on ${\cal M}$ related by ${\cal G}$ are considered equivalent, and a fundamental
domain for the moduli space arises from dividing out by ${\cal G}$.

For the vacua we consider the fluxes are affected by ${\cal G}$ as well.
$SL(2,Z)_\tau$ acts in the ordinary way: given an $SL(2,Z)$ matrix
$\pmatrix{a \;\;\;b \cr c \;\;\; d}$ we have
\eqn\sltwoz{
\tau \rightarrow {a \, \tau + b \over c\,  \tau + d}   \,, \quad \quad
\pmatrix{f \cr h} \rightarrow \pmatrix{a \;\;\; b \cr c \;\;\; d} \pmatrix{f \cr h} \,.
} Under this transformation $(f - \tau h) \rightarrow (f - \tau h)/(c
\tau + d)$, hence solutions of \fflat\ are carried into other
solutions, and $N_{\rm flux}$ \nflux\ is preserved.  The action of
$SL(2,Z)$ generates a K\"ahler transformation on $W$ \gvw\ and ${\cal K}$ \kahler:
\eqn\kahlertrans{
W \rightarrow \Lambda W \,, \quad \quad {\cal K} \rightarrow {\cal K}
- \log \Lambda - \log \bar\Lambda \,,
}
with in this case $\Lambda = 1 /(c \tau + d)$.

\newsec{Supersymmetric flux vacua as New Attractors}

The relation of flux vacua notation in the previous section
to the notation used in section \S2 for New Attractors is as follows.
The covariantly-holomorphic section is related to the holomorphic one as
\eqn\rel{
\Pi \rightarrow e^{{\cal K }\over 2}\Pi(z)
}

The fluxes $f$ and $h$ of section \S2
are related to fluxes of section \S3 as follows
\eqn\relflux{
f \rightarrow \Sigma\cdot f \qquad\qquad
h \rightarrow \Sigma\cdot h
}

Finally the relation of the superpotential is
\eqn\suprel{
Z = e^{{\cal K} \over 2}W
}

New Attractors \atreq\ can now be rewritten in terms of flux vacua notations as
\eqn\newattractors{
\pmatrix{\Sigma\cdot h \cr \Sigma \cdot f}
= e^{\cal K}\left.\pmatrix{ 2{\rm Re}(W{\overline \Pi}) \cr 2{\rm Re}(\bar{\tau}W {\overline \Pi})}
\right|_{DW=0}
+ e^{\cal K}\left.\pmatrix{ 2{\rm Re}(e^{\tau}_{\underline 0}e^{\psi}_{1}e^{\overline \psi}_{\overline 1}D_{\psi}D_{\tau}W {\overline {D_{\psi}\Pi}})
\cr 2{\rm Re}(e^{\tau}_{\underline 0}e^{\psi}_{1}e^{\overline \psi}_{\overline 1}\tau D_{\psi}D_{\tau}W{\overline {D_{\psi}\Pi}}) \cr }
\right|_{DW=0}
}
where for simplicity we assume that we have only one complex structure modulus $\psi$
(as will be the case in examples in the remaining part of this note) and
\eqn\ee{
e^{\psi}_{1}e^{\overline \psi}_{\overline 1} =
g^{\psi\bar{\psi}} = {1\over \partial_{\psi}\partial_{\bar{\psi}}{\cal K}}
\qquad\qquad
e^{\tau}_{\underline 0} = -(\tau-\bar\tau)
}

\newsec{Flux vacua as New Attractors in a simple Calabi-Yau hypersurface}

Consider a Calabi-Yau threefold defined as a hypersurface
in a weighted projective space.
The Calabi-Yau threefold of interest is defined by the equation
\eqn\model{
\sum_{i = 1}^{4}x_{i}^{6}+ 2x_0^3 - 6\psi\,x_0 x_1 x_2 x_3 x_4 = 0 \qquad  x_i \in WP^{4}_{1,1,1,1,2}
}
On its moduli space it has Landau-Ginzburg, conifold and large complex structure points.
We will analyze the flux vacua at the Landau-Ginzburg point and
in vicinity of the conifold point. Both of these points are at a finite distance in moduli space.

\subsec{Flux vacua at Landau-Ginzburg point}

In general the Landau-Ginzburg point is a very special point in the
moduli space, where the number of vacua with $W = 0$ and with discrete
symmetries can be of the same order as the total number of vacua, when nonzero \DGKT.

 Near the Landau-Ginzburg point $\psi = 0$
the periods admit expansion in a Pichard-Fuchs basis
\eqn\periods{
w_{i}(\psi) = {(2\pi i)^{3}\over 6}\sum_{n=1}^{\infty}{\exp({5\pi i\over 6}n)\Gamma({n\over 6})
\over \Gamma(n)\Gamma(1-{n\over 6})^3\Gamma(1-{n\over 3})}
\left({6\alpha^{i}\over 2^{1/3}}\right)^{n}\psi^{n-1}.
}
This is valid for $|\psi| < 1$, where $\alpha$ is the $6^{\rm th}$ root of unity
\eqn\alphadef{\alpha = \exp\left({2\pi i\over 6}\right).}

In symplectic basis the periods  then have the expansion
\eqn\per{
\Pi = \pmatrix{g_1 \cr g_2 \cr z^1 \cr z^2 \cr} = m\cdot \pmatrix{w_2 \cr w_1 \cr w_0 \cr w_5 \cr}
= c_0 p_0 + c_1 \psi p_1 + O(|\psi|^2)
}
around the LG point $\psi$. Here $c_0$, $c_1$ are constants
and the matrix of transformation from the Pichard-Fuchs to the symplectic basis is given by
\eqn\mmatrix{
m = \pmatrix{-{1 \over 3} & -{1 \over 3} & {1 \over 3}
& {1 \over 3}\cr 0 & 0 & -1 & 0\cr -1 & 0 & 3 & 2 \cr 0 & 1 & -1 & 0\cr}
}
and the following definitions are introduced
\eqn\pvectos{
p_0 = \pmatrix{\alpha^2 \cr \alpha \cr 1 \cr \alpha^5 \cr}
\qquad\qquad
p_1 = \pmatrix{\alpha^4 \cr \alpha^2 \cr 1 \cr \alpha^4 \cr}.
}
The monodromy group, $\Gamma$, of the complex structure moduli space has two generators:
$A$, which generates phase rotations $\psi \to \alpha \psi$ with $\alpha = \exp (2 \pi i /6)$
around the LG point at $\psi = 0$, and $T$ which corresponds to the logarithmic monodromy
${\cal G}_2 \to {\cal G}_2 + z^2$ around the conifold singularity $\psi = 1$.  By itself, $A$
generates a ${\bf Z}_6 \subset \Gamma$ subgroup, with an associated fixed point at $\psi = 0$;
$T$, on the other hand, is of infinite order.

F-flatness condition at $\psi = 0$ reduces to
\eqn\dtau{
D_{\tau}W = (f - \bar{\tau}h)\cdot p_0 = 0
}
\eqn\dpsi{
D_{\psi}W = (f - \tau h)\cdot p_1 = 0.
}

The monodromy matrix $A$  generates rotations by a root of unity around $\psi =0$:
\eqn\sixaperiods{
A \Pi(\psi) = \alpha \Pi(\alpha \psi) \,,
}
and is explicitly given by
\eqn\asix{A =\pmatrix{1 & -1 & 0 & 1\cr
0 & 1 & 0 & -1\cr -3 & -3 & 1 & 3 \cr -6 & 4 & 1 & -3\cr}.}

Vacua exist at the LG point as long as
\eqn\taucond{\tau = t_1 +\alpha t_2}
where $t_1$ and $t_2$ are rational \DGKT.

The LG point is a fixed point for ${\bf Z}_6 \subset \Gamma$, so one may
hope that this symmetry is preserved in the low-energy theory.
Additionally, the ${\bf Z}_2$ and ${\bf Z}_3$ points on the dilaton moduli space
are also potential sources of low-energy symmetry.  Only ${\bf Z}_3$ is accessible
though.

\subsec{$(0, 3)$ flux vacua}

In addition to F-flatness these flux vacua satisfy the condition
\eqn\threezerocond{D_{\tau}D_{\psi}W=0}
which results in the following constraint for fluxes which define the vacuum
\eqn\condddwzero{
f\cdot p_1 = h\cdot p_1 = 0
}
This is solved by the following
\eqn\ddwzerosolh{
h = -h\cdot A^{3} \qquad\rightarrow\qquad h = (-3h_3 + 3h_4, h_3, h_3, h_4)
}
\eqn\ddwzerosolf{
f = -f\cdot A^{3} \qquad\rightarrow\qquad f = (-3f_3 + 3f_4, f_3, f_3, f_4).
}
Hence in this case we have ${\bf Z}_2 \subset {\bf Z}_6$ preserved as a true symmetry
(see \DGKT\ for details):
\eqn\wsixeven{
\quad W(\tau, -\psi) = W(\tau, \psi).
}

Let us now see if the New Attractor equations hold for these flux vacua.
The second term in New Attractors \newattractors\ vanishes and they become
\eqn\oldattr{
\pmatrix{\Sigma\cdot h \cr \Sigma \cdot f}
= e^{\cal K}\left.\pmatrix{ 2{\rm Re}(W{\overline \Pi}) \cr 2{\rm Re}(\bar{\tau}W {\overline \Pi})}
\right|_{\matrix{\psi = 0 \cr \tau = {f\cdot p_0^{\dagger}\over h\cdot p_0^{\dagger}}}}
}
We checked this using computer algebra as we do most of the checks in this note.

\subsec{$(2, 1)$ flux vacua}

In addition to F-flatness these vacua satisfy the condition $W = 0$
which results in the following necessary and sufficient condition
\eqn\condwzero{
f\cdot p_0 = h\cdot p_0 = 0.
}
This is solved by the following choice of fluxes
\eqn\wzerosolh{
h = h\cdot A^{3} \qquad\rightarrow\qquad h = (-3h_3 + h_4, 3h_3, h_3, h_4).
}
\eqn\wzerosolf{
f = f\cdot A^{3} \qquad\rightarrow\qquad f = (-3f_3 + f_4, 3f_3, f_3, f_4)
}
Hence these vacua have a ${\bf Z}_2$ R-symmetry (see \DGKT\ for details)
\eqn\wsixodd{
W(\tau, -\psi) = - W(\tau, \psi)
}
and it is easy to see that this ensures $W (\tau, \psi = 0) = 0$.
This is the R-symmetry ``responsible"
for the vanishing of the vacuum superpotential.

In checking the New Attractor equations we see that
the first term now vanishes and they become
\eqn\nnnnattr{
\pmatrix{\Sigma\cdot h \cr \Sigma \cdot f}
= e^{\cal K}\left.\pmatrix{ 2{\rm Re}(g^{\psi\bar{\psi}}e^{\tau}_{\underline 0}D_{\psi}D_{\tau}W {\overline {D_{\psi}\Pi}})
\cr 2{\rm Re}(g^{\psi\bar{\psi}}e^{\tau}_{\underline 0}\tau D_{\psi}D_{\tau}W{\overline {D_{\psi}\Pi}}) \cr }
\right|_{\matrix{\psi = 0 \cr \tau = {f\cdot p_1\over h\cdot p_1}}}
}
where the metric at Landau-Ginzburg point on complex structure moduli space is given by
\eqn\eee{
g^{\psi\bar{\psi}} = {|c_1|^2\over 3|c_0|^2}.
}
Using computer algebra it is easy to check that these equations are indeed satisfied.

\subsec{$(0, 3) + (2, 1)$ flux vacua}

Generic flux vacua (with $W$ and $D_{\tau}D_{\psi}W$ not equal zero
generically) are defined by the condition
\eqn\gencondfh{
f = t_1 h - t_2 h\cdot A^2
}
for any integral $h$ and rational $t_1$, $t_2$ chosen so that $f$
is integral and the axio-dilaton is given by
\eqn\taueq{\tau = t_1 + \alpha t_2.}

In this case the New Attractors \newattractors\ should work as well , once again
using computer algebra, it is easy to check that they
are satisfied.

\subsec{Flux vacua at conifold region}

Let us now study supersymmetric flux vacua in conifold region of the $WP^{4}_{1,1,1,1,2}$
model.

In a symplectic basis the periods in the vicinity of the conifold point $\psi=1$
can be given to first order by the following expressions
(here $x\equiv 1-\psi$ and $|x|\ll 1$)
\eqn\perconi{\eqalign{
& \CG_1(x)=(2\pi i)^3[a_0+a_1x+O(x^2)],\cr
& \CG_2(x)={z^2(x)\over 2\pi i}\ln(x)+(2\pi i)^3[b_0+b_1x+O(x^2)],\cr
& z^1(x)=(2\pi i)^3[c_0+c_1 x+O(x^2)],\cr
& z^2(x)=(2\pi i)^3[d_0+d_1 x+O(x^2)].
}}
Where the constants can be approximated by the
following numbers
\eqn\perconia{\eqalign{
&a_0=1.501 i,\qquad\qquad\qquad\; c_0=-5.087 + 6.754 i, \cr
&a_1=-0.914 i, \qquad\qquad\;\;\;\;\;\; c_1=4.261 -4.112 i,\cr
&b_0=1.056, \qquad\qquad\qquad\;\;\; d_0=0,\cr
&b_1=-0.344-0.827 i,\quad\quad~ d_1=-1.654 i.
}}

The K\"ahler potential for the complex structure modulus
is given by \eqn\Kconi{ K_{\psi}=-\ln[
\mu_0+\mu_1 x + \bar{\mu}_1\bar{x} +\mu_2 |x|^2\ln|x|^2
+\mu_3 |x|^2+\mu_4 x^2+\bar{\mu}_4 \bar{x}^2+O(|x|^3\ln|x|)],
} with the relevant constants $\mu_0$, $\mu_1$, $\mu_2$ and $\mu_3$
given by
\eqn\Kconia{\eqalign{
& \mu_0=i(2\pi)^6(a_0 \bar{c}_0-c_0 \bar{a}_0),\qquad
 \mu_1=i(2\pi)^6(\bar{c}_0 a_1-c_1 \bar{a}_0-d_1\bar{b}_0),\cr
& \mu_2=(2\pi)^5|d_1|^2,\qquad
  ~~~~~~~~~~~~
  \mu_3=i(2\pi)^6(\bar{c}_1 a_1-\bar{a}_1 c_1+\bar{d}_1 b_1-\bar{b}_1 d_1).
}}
One finds the following expression for the K\"ahler metric
\eqn\metrconi{
g_{x\bar{x}}=-{\mu_2\over\mu_0}\ln|x|^2
+\left({|\mu_1|^2\over \mu_0^2}-{2\mu_2+\mu_3\over \mu_0}\right)+O(|x|\ln|x|).
}
In computing K\"ahler covariantized derivatives with respect to
$\psi$ it is also useful to note that
\eqn\kahldirconi{
\partial_{x}K_{\psi}=-{\mu_1\over \mu_0}-{\mu_2\over\mu_0}\bar{x}\ln|x|^2+O(x).
}

\subsec{Flux vacua as New Attractors at the conifold point}

The approximate equations for supersymmetric flux vacua that are
very close to conifold point take the form
\GKT
\eqn\condconia{
D_{\tau}W=0\quad\Rightarrow\quad
\tau={f\cdot \Pi^{\dagger}\over h\cdot \Pi^{\dagger}}=
{f_1\bar{a}_0+f_2\bar{b}_0+f_3\bar{c}_0\over h_1\bar{a}_0+h_2\bar{b}_0+h_3\bar{c}_0}
+O(|x|\ln|x|);
}
\eqn\condconib{\eqalign{
& D_{\psi}W=0\quad\Rightarrow\quad
 \ln(x)=-{2\pi i\over d_1}\left[
{(f_1-\tau h_1)(a_1-{\mu_1\over \mu_0}a_0)
+(f_2-\tau h_2)(b_1-{\mu_1\over \mu_0}b_0)\over f_2-\tau h_2}+\right.\cr
&\left.\qquad\qquad\qquad\qquad\qquad\qquad
\qquad\qquad +{(f_3-\tau h_3)(c_1-{\mu_1\over \mu_0}c_0)
+(f_4-\tau h_4)d_1
\over f_2-\tau h_2}\right]-1.
}}

One can make Monte Carlo simulations of such vacua. This was done
in \GKT\ where the attractive nature of conifold point was established.
We will take one particular choice of fluxes to illustrate that the
New Attractors \newattractors\ work.
For the choice of fluxes
\eqn\fluxesconi{
f = \{61, 10, -11, -15\}
\qquad\qquad
h = \{3, 4, 4, 11\}
}
and using \condconia, \condconib\ , one finds a supersymmetric vacuum deep in the
conifold region with
\eqn\vacconi{
\ln (1-\psi) = -7.29 - 0.71 i
\qquad\qquad
\tau = -0.42 + 1.90 i
}
One may check that the F-flatness conditions
\eqn\fflatncodn{
D_{\tau}W \approx 0
\qquad\qquad
D_{\psi}W \approx 0
}
hold with precision $O(10^{-2})$ and New Attractors hold with precision $O(10^{-3})$.

\newsec{Area codes for flux vacua}

A new phenomena that we observe in this note is that certain fluxes may give rise to multiple minima
within a finite distance on the moduli space. This suggests the existence of area codes and basins
of attraction for these flux vacua.

We give an illustration of these multiple basin attractors here.
For simplicity let us consider flux vacua with the ${\bf Z}_3$ symmetry
on the dilaton moduli space being preserved in the low-energy action.
In order for fluxes to permit $\tau = \alpha$ as a vacuum we must restrict
the Landau-Ginzburg solutions \DGKT\ to fluxes satisfying
\eqn\zthreesoln{
f = - h \cdot A^2.
}

To be specific let us take the $H_{(3)}$ flux to be
\eqn\hfluxchoice{
h = \{-72, -3, 13, -4\}
}
Note this supersymmetric flux vacuum with enhanced symmetry satisfies the New Attractors
\newattractors\ of course.

It turns out, using \condconia\ and \condconib\ , that this flux give rise to a
vacuum deep in conifold region
with
\eqn\psitau{
\tau = 0.37 + 1.11 i
\qquad\qquad
\ln(1-\psi) = -18.47 - 0.37 i
}
Computer algebra confirms that the F-flatness conditions
and the New Attractor equations hold with precision $O(10^{-3})$.

This means that for flux choice \hfluxchoice\ there are at least
two supersymmetric AdS flux vacua,
with cosmological constant $V = -3e^{\cal K}|W|^2$ given by
\eqn\twopot{
V_{\rm LG} = -2080.5
\qquad\qquad
V_{\rm conifold} \simeq -2281.7
}
which are not the same but are close.

This suggests the possible existence of basins of attraction
and domain walls in the landscape domain walls between two
flux vacua which arise for a fixed flux and both of which are at interior
points in moduli space.

\newsec{Non-supersymmetric extremal black holes}

In this section we switch gears and consider the non-supersymmetric extremal
black holes proposed in \FGK\ and rediscovered and developed more recently
in \Trivedi, \Sen\ and \TripathyTrivedi.

Let us consider the same $WP^{4}_{1,1,1,1,2}$ model and construct
non-supersymmetric extremal black hole attractors at the Landau-Ginzburg
point of this model.

In ${\cal N}=2$ supersymmetric theory the effective black hole potential
can be expressed as
\eqn\veff{
V_{\rm BH} = e^{\cal K}(|DW|^2+|W|^2)
}
in terms of a K\"ahler potential given by
\eqn\kahlerblack{
{\cal K} = - \log(-i \Pi^\dagger \cdot \Sigma \cdot \Pi)
}
and a superpotential which reads as
\eqn\supblack{
W = \int_{M}\, F_{(3)}\wedge \Omega = f\cdot \Pi
}
where $f$ is the $F_{(3)}$ magnetic and electric flux.

Non-supersymmetric black hole attractors are just critical points of effective black hole potential \FGK
\eqn\potmin{
\partial_{i}V_{\rm BH} = 0
}
and subject to attractors \Trivedi\ if
\eqn\potmin{
M_{ij}={1\over 2}\partial_{i}\partial_{j}V_{\rm BH} > 0.
}

In this case the resulting black hall entropy is just given by
\eqn\entropy{
S_{\rm BH} = \pi V_{\rm BH}
}
at the minimum with a positive definite mass matrix.

Let us now go back to our model with one complex structure modulus.
It turns out that this minimum of the effective black hole potential
\eqn\veffextr{
V_{\rm eff}(\psi) = e^{\cal K}(G^{\psi\bar\psi}|D_{\psi}W| + |W|^2)
\qquad\qquad
\partial_{\psi}V_{\rm eff}(\psi) = 0
}
happens to be at Landau-Ginzburg point for fluxes
\eqn\wzerosolfff{
f = (-3f_3 + f_4, 3f_3, f_3, f_4).
}
This is equivalent to $f = f\cdot A^{3}$ in terms of Landau-Ginzburg monodromy matrix $A$
defined in \asix.

The mass matrix for these fluxes is found to be
\eqn\mass{
\partial_{\psi}\partial_{\psi}V_{\rm BH}(\psi)|_{\psi=0} = 0\qquad\qquad
\partial_{\psi}\partial_{\bar\psi}V_{\rm BH}(\psi)|_{\psi=0} = 2\,
 G_{\psi\bar\psi}|_{\psi=0}\,V_{\rm BH}|_{\psi = 0}
}
where the effective black hole potential at the attractor point is
\eqn\pot{
V_{\rm BH}(0) = {2\over\sqrt{3}}(3f_3^2+3f_3 f_4 + f_4^2).
}
This means that these critical points are indeed attractors for all possible fluxes \wzerosolfff.
Note also that for this case we have $W = 0$, though this is non-generic as we
see from the examples in \TripathyTrivedi.

\subsec{New Attractors for non-supersymetric extremal black holes}

Let us finally check that the recently proposed New Attractor equations for non-supersymmetric
extremal black holes \Kalloshtwo\ holds. This equation is
\eqn\nonsusyattr{
f = 2e^{\cal K}{\rm Im}(W{\overline \Pi} - G^{a\bar{a}}D_{a}W{\overline {D_{a}\Pi}}).
}
In our simple case of a one dimensional moduli space (and our particula choice of fluxes) these
equations become
\eqn\nonsusyattrour{
\Sigma\cdot f = 2e^{\cal K}{\rm Im}(W{\overline \Pi} - G^{\psi\bar{\psi}}D_{\psi}W{\overline {D_{\psi}\Pi}})
}
It is a matter of simple algebra to check that these equations
are indeed satisfied for the solutions above.

We finish our note with the remark that through the explicit examples in the sections above
we have illustrated phenomena which certainly deserve further attention.

\newsec{Conclusion}

  In this note we have found multiple examples of supersymmetric flux vacua which satisfy
new supersymmetric attractor equations. We have also found examples of non-supersymmetric
black hole attractors which solve the corresponding non-supersymmetric New Attractor equations.
There is only one class of new non-supersymmetric  attractor equations for flux vacua
for which we do not check the new equations. It would be interesting to do it.

 Though the existence of domain walls of various sorts in the landscape was
already known; e.g. the domain wall between the KKLT vacuum and infinity,
or the domain walls which jump the quantized RR and NS flux and are given
by D and NS branes wrapping 3-cycles. In this note however we are giving
the first signs of existence of domain wall between two flux vacua which
arise for a fixed flux and both of which are at interior points in moduli space.

With this regard we should say that a first bottom-up pass at the "measure problem"
for vacuum selection would clearly involve doing statistics of the sizes of basins
of attraction (in the metric on CY moduli space) for the various vacua which arise at a fixed
flux.

$\;$

$\;$

\centerline{\bf{Acknowledgements}}

I would like to thank R. Kallosh for motivating this work and N.
Sivanandam for a significant help. I benefited greatly from
interesting discussions with D. Belov, F. Denef, O. DeWolfe, B.
Florea, S. Kachru, X. Liu, E. Silverstein, W. Taylor, P. Tripathy
and S. Trivedi. The work was partially supported by NSF grant
PHY-0097915, the DOE under contract DE-AC03-76SF00515
and RFBR 05-01-00758.

\listrefs

\end